\documentclass[amsmath,amssymb,aps,twocolumn,showpacs,longbibliography,floatfix,prl,lengthcheck]{
revtex4}
\usepackage{graphicx}
\usepackage{bm}
\usepackage{hyperref}

\begin{document}
\title{Kinetics of a frictional granular motor}
\author{J. Talbot$^1$, R. D. Wildman$^{1,2}$ and P. Viot$^1$} 
\affiliation{$^1$ Laboratoire de Physique
  Th\'eorique de la Mati\`ere Condens\'ee, UPMC, CNRS  UMR 7600, 4, place Jussieu, 75252 Paris Cedex 05, France}
\affiliation{$^2$ School of Mechanical and Manufacturing Engineering, Loughborough University,
Loughborough, Leicestershire, LE11 3TU, UK}
\date{\today}
\begin{abstract} 
Within the framework of  a  Boltzmann-Lorentz equation, we analyze the dynamics
of a  granular rotor immersed in a bath of thermalized particles in the presence
of a frictional torque on the axis. In numerical simulations of the equation, we observe two scaling regimes 
at low and high bath temperatures. In the large friction
limit, we obtain the exact solution of a model corresponding to asymptotic
behavior of the Boltzmann-Lorentz equation. In the limit of large rotor mass  and small friction, we derive a Fokker-Planck 
equation for which the exact solution is also obtained. 
\end{abstract}
\pacs{45.70.-n,45.70.Vn,05.10.Gg}
\maketitle

Recently, Eshuis {\it et al}. \cite{PhysRevLett.104.248001}, inspired by Smoluchowski's gedankenexperiment,  
constructed a macroscopic rotational motor
consisting of four vanes immersed in a granular gas. When a soft coating was
applied to one side of each vane, 
a ratchet 
effect was observed above a critical granular temperature. 
While this was the first experimental realization of a granular motor, similar Brownian ratchets exist in many diverse applications, e.g., photovoltaic devices and biological motors; See \cite{Reimann2002}.
All of these motors share the common features of non-equilibrium conditions and spatial symmetry breaking. 
Several recent theoretical studies of idealized models of granular motors, which use a Boltzmann-Lorentz description
\cite{Cleuren2007,costantini:061124,Cleuren2008,Costantini2008,Costantini2009,PhysRevE.82.011135}, show that the motor effect is particularly pronounced
when  the device is constructed from two different materials, as was the case in the recent experiment \cite{PhysRevLett.104.248001}.
The existing theories, however, predict a motor effect for any temperature of the granular gas while in the
experiment the phenomenon is only observed if the bath temperature is sufficiently large. The presence of friction is at the origin of this difference and it is therefore highly desirable to incorporate 
it in the theoretical schemes.

Several studies have considered the effect of friction on Brownian motion. The pioneering work of de Gennes \cite{Gennes2005} was motivated
by the motion of a coin on a horizontally vibrating plate\cite{A.Buguin2006} and a liquid droplet on non-wettable surfaces
subjected to an asymmetric lateral vibration; 
he obtained the stationary velocity distribution, as well as the velocity correlation function. 
A notable result, obtained independently by
Hayakawa \cite{Hayakawa2005}, is that the Einstein relation no longer holds in the presence of dry friction. 
More recently, Touchette and coworkers
\cite{Touchette2010,Baule2010,Baule2011}  obtained a solution of a model with dry friction and viscous damping.

In this Letter, we  investigate the
kinetic properties of a heterogeneous granular rotor in the presence of dry (Coulomb) friction and 
we provide exact solutions in the limit of large motor mass (Brownian limit) and also in the limit 
of strong friction for arbitrary mass . 

We  consider a two-dimensional system where a  heterogeneous chiral rotor with
moment of inertia $I$, mass $M$ and the length $L$ immersed
in a granular gas at density $\rho$ 
with a velocity distribution $\phi(v)$ characterized by a granular temperature $T$. 
The motor is composed of two materials with 
coefficients of restitution $\alpha_{+}$ and $\alpha_{-}$. Collisions of
the bath particles, of mass $m$, 
with the former (latter) exert a positive (negative) torque. (The experimental
setup employed by Eshuis et al., where one has a translational invariance in the third
dimension, can also be described by the model).  

\begin{figure}[h]
\begin{center}
\resizebox{3.5cm}{!}{\includegraphics{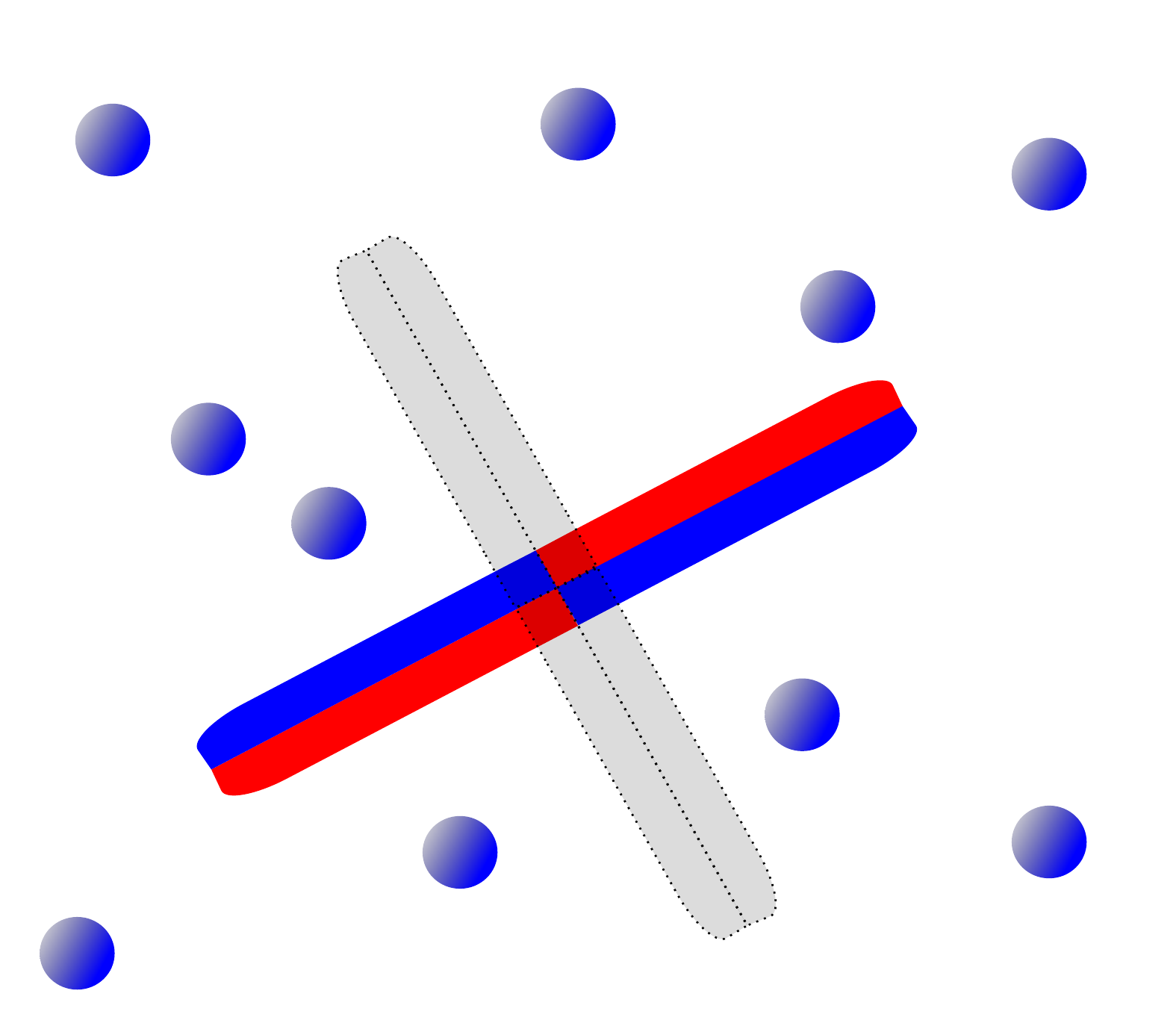}}
\end{center}
\caption{
  The chiral rotor is constructed of materials with
  coefficients of restitution $\alpha_{+}$ (red) and $\alpha_{-}$ (blue). When 
immersed in a bath of thermalized particles, it acquires a net
  rotation around its axis. The model can be extended to include additional vanes 
(shown in gray) as used in the experiment of Eshuis et al. \cite{PhysRevLett.104.248001}.}
\label{fig:rotor}
\end{figure}

To understand the qualitative effect of the friction, it is useful to consider the torques acting on the rotor: The motor
torque is  the average of all  changes of angular momentum resulting from collisions
between the rotor and the bath particles. In the large rotor mass ratio ($M/m>>1$), one finds
that the motor torque is
$(\alpha_+ -\alpha_-)\frac{1}{8}\rho L^2 T$ \cite{Talbot2011}, 
which vanishes for a symmetric rotor ($\alpha_+=\alpha_-$) and the dynamic friction is
assumed constant and opposed to the motion $-\Gamma_s\sigma(\Omega)$ ($\sigma(\pm x)=\pm 1$, $|x|\neq 0$; $\sigma(0)=0$).
(We neglect static friction  in this study, but it will be considered in a future publication).
We can define a crossover temperature, 
$T_c=\frac{8\Gamma_s}{\rho L^2(\alpha_+-\alpha_-)}$, that separates a regime dominated by the motor effect 
($T>T_c$) and one dominated by 
friction ($T<T_c$).  This mechanical point of view defining the two kinetic regimes must be expanded by accounting for the fact that 
the motor torque is a 
fluctuating quantity.

Assuming the bath is homogeneous, and in
the presence of dry friction,   the Boltzmann-Lorentz (BL) equation of the angular velocity distribution (AVD), 
$f(\Omega)$, can be  
expressed as
\begin{align}\label{eq:BL}
&\frac{\partial}{\partial t} f(\Omega;t) -\frac{\Gamma_s\sigma(\Omega)}{I}\frac{\partial}{\partial \Omega} f(\Omega;t)= 
J[f,\phi]
\end{align}
where the frictional torque  $\Gamma_s$   and 
$J[f,\phi]$ is the BL collision operator \cite{Talbot2011}
\begin{align}&J[f,\phi]=
\rho \int_{-\infty}^{\infty} dv \,\int_{-L/2}^{L/2} d\lambda
|v-\lambda\Omega| [\theta(v-\lambda\Omega) \frac{f(\Omega^{**} ;t)}{\alpha_{+}^2}
\nonumber\\&\phi (v^{**})
 + \theta(\lambda\Omega-v) \frac{f(\Omega^{**} ;t)}{\alpha_{-}^2}\phi (v^{**}) ] - \rho
\nu(\Omega)f(\Omega;t)
\end{align} 
with $\Omega^{**},v^{**}$ denoting the pre-collisional velocities, $\theta(x)$ is the Heaviside function and $\rho \nu(\Omega)$ is the total collision rate  for a 
motor rotating at the angular velocity $\Omega$, $
\rho\nu(\Omega)=\rho\int_{-\infty}^{\infty} dv \int_{-L/2}^{L/2} d\lambda\, |v-\lambda\Omega|\phi (v)$. (For a $3D$-motor, Eq.\ref{eq:BL}
is unchanged if $\rho=nL_1$, $n$ being the $3D$ number density and $L_1$ the perpendicular linear size of the vane.)
For convenience we introduce the dimensionless variables
$\Omega^*=L\sqrt{\frac{m}{T}}\Omega$, $\Gamma_s^*=\frac{\Gamma_s}{\rho L^2T}$ and $v^*=v\sqrt\frac{m}{T}$. 
With this choice of units, the steady state angular velocity depends only on $\alpha_{\pm}$, $M/m$ and $\Gamma_s^*$.
Also, since the BL description  neglects recollisions with the bath particles and is invalid when $M<<m$, we will assume in the following that $M>m$.

{\em Simulations.} A complete analytical treatment of the Boltzmann equation is a very difficult
task, even in the steady state. 
\begin{figure}[t]
\resizebox{7cm}{!}{\includegraphics{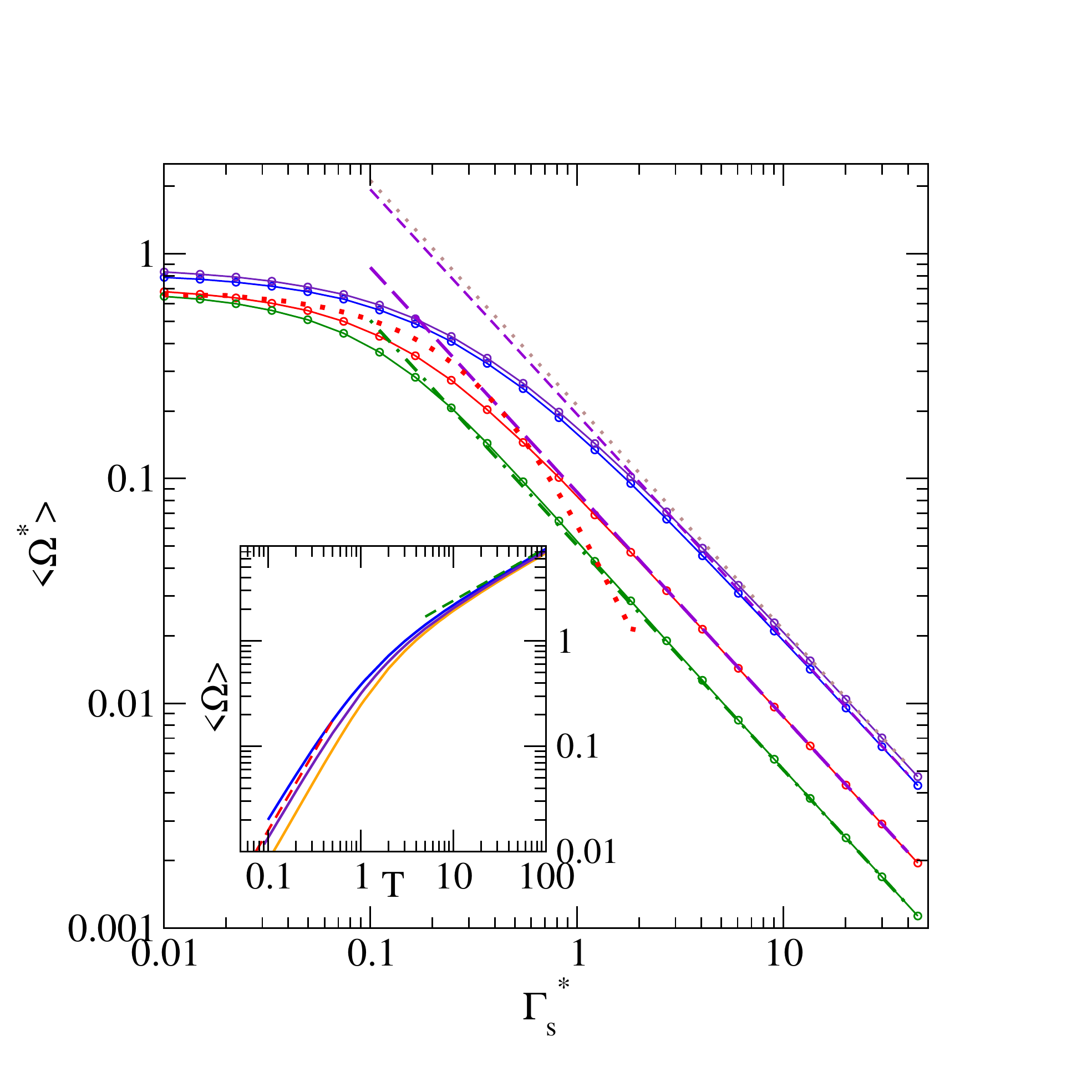}}
 \caption{(Color online) Dimensionless mean angular velocity $\langle \Omega^* \rangle$ of 
an asymmetric  rotor versus $\Gamma_s^*$:  $\alpha_{+}=1$, $\alpha_{-}=0$ for different mass ratios 
$M/m=1,2,10,20$, top to bottom obtained from numerical simulations of the BL equation.  The dashed and dotted curves correspond to the solutions of
the Independent Kick model and the
Fokker-Planck equation, respectively. The inset shows  $\langle \Omega\rangle$ as a function of the bath granular 
temperature, $T$, 
 for $\Gamma_s=0.2$ and $M/m=5,10,20$. At low temperature $<\Omega>\sim T^{3/2}$ and at high temperature $<\Omega>\sim T^{1/2}$. 
The dashed lines are visual guidance for this scaling behavior.}
 \label{fig:velocity_fric6}
 \end{figure}
We have performed numerical simulations of the
BL equation. (See \cite{TV06} for details. 
We note that both in simulation and experiment the sampling time must be smaller than the mean stopping time 
to avoid biasing the AVD). Fig. \ref{fig:velocity_fric6} shows the dimensionless mean angular velocity $\Omega^*$ of an asymmetric
 rotor ($\alpha_{+}=1$,
$\alpha_{-}=0$) as a function of a dimensionless
dry friction $\Gamma_s^*$ for different mass ratios $M/m=1,2,10,20$.
As expected, the mean angular velocity  decreases as the friction increases. Moreover, two scaling regimes are observed:
\begin{equation}
 \langle\Omega^*\rangle\sim \left\{\begin{array}{cc}
                                    \Gamma_s^{*\;-1} ,&\Gamma_s^{*}>>1\\
                                    \Gamma_s^{* \;0}, &\Gamma_s^{*}<<1
                                    \end{array}
\right.
\end{equation}

In experiments, the most easily varied parameter is the bath temperature $T$, and
in terms of the original variables, the 
 two  scaling regimes correspond to
$\langle\Omega\rangle \propto T^{3/2}$ and $\langle\Omega\rangle\sim T^{1/2}$ at low and high
temperatures, respectively. (The inset of Fig.~\ref{fig:velocity_fric6} shows a 
log-log plot of the mean angular velocity $\langle\Omega\rangle$ as a function of bath temperature $T$.).

{\em Limiting Behaviors}. We now show that
in appropriate limits, the physics can be described by simpler models for which
analytical solutions can be obtained. The analysis is facilitated 
by considering two
time-scales: 
the average inter-collision time between bath particles and the motor,
$\tau_{c}\simeq\frac{1}{\rho\nu(0)}=\sqrt{\frac{\pi m}{2T}} \frac{1}{\rho L}$ (for a Gaussian bath distribution)
and the mean stopping time in the presence of friction, 
$ \tau_{s}=\frac{I\overline{\Omega}}{\Gamma_s}$
where $\overline{\Omega}$ is the average angular velocity after a collision.
When the dimensionless friction $\Gamma_s^*$ is small, $\overline{\Omega^*}\sim (\alpha_{+}-\alpha_{-})$, while for $\Gamma_s^*>>1$, 
$\overline{\Omega^*}\sim \frac{mL^2}{I}(\alpha_{+}-\alpha_{-})$, which gives
\begin{equation}
 \frac{\tau_s}{\tau_c}\sim \frac{\alpha_{+}-\alpha_{-}}{\Gamma_s ^*}\left\{\begin{array}{cc}
                                    1 ,&\Gamma_s^{*}>>1\\
                                    \frac{M}{m}, &\Gamma_s^{*}<<1
                                    \end{array}
\right.
\end{equation}

Whenever slip-stick behavior is present the  AVD 
contains a regular part and a delta singularity at $\Omega^*=0$ corresponding to a stationary rotor before the 
next collision with a bath particle: 
\begin{equation}
 f^*(\Omega^*)=\gamma f_R^*(\Omega^*)+(1-\gamma)\delta(\Omega^*)
\end{equation}
where $\int d\Omega^* f_R^*(\Omega^*)=1$ and $\gamma$ is a constant that can be determined from conservation of the probability current 
at $\Omega^*=0$\cite{Touchette2010}:
\begin{equation}
(1-\gamma)\rho L\int_{-\infty}^{\infty} dv^* |v^*| \phi(v^*)=2\gamma
f_R^*(0)\frac{\Gamma_s^*mL^2}{I}
\end{equation}
giving $\gamma^{-1}=1+2C \frac{\Gamma_s^* mL^2}{I}$ with $C=2f_R^*(0)/\int_{-\infty}^{\infty} dy |y| \phi(y)$ is a numerical constant

When $\tau_c>>\tau_s$ the dry friction 
stops the motor before the next collision, 
and the motor essentially evolves by following a sequence of stick-slip
motions. Most of the time, the rotor is at rest and the singular
contribution is dominant, $\gamma\simeq 1/\Gamma_s^*$. This regime can be
described by the Independent Kick model introduced below. 

Conversely, when $\tau_c<<\tau_s$, collisions are so 
frequent that sliding dominates the motor dynamics. For all practical purposes, the rotor
never stops and $(1-\gamma)\simeq \Gamma_s^*$.
In this case, the dynamics is well described by 
a Fokker-Planck equation for $M/m>>1$.

{\em Independent Kick Model}. The mean angular velocity is the average over all collisions:
\begin{equation}
\langle \Omega \rangle =\rho \int_{-\infty}^{\infty} dv |v| \phi(v)
\int_{-L/2}^{L/2}dx \int_0^\tau \Omega(t)dt,
\end{equation}
where $\Omega(t)=\Omega_0-\frac{\Gamma_s\sigma(\Omega_0)}{I}t$, $\tau=\frac{|\Omega_0|
I}{\Gamma_s}$ and $\Omega_0$ is the angular velocity after a collision, which
is given by
\begin{equation}
 \Omega_0=\frac{(1+\alpha_{\pm})mxv}{I+mx^2}
\end{equation}
Integrating over time and $x$, and by using a Gaussian bath 
distribution (analytical expressions can be obtained for arbitrary bath 
distributions, but they are not presented here), the dimensionless mean angular velocity is given by
\begin{equation}\label{eq:omega_asym}
\langle \Omega^* \rangle
=\frac{(1+\alpha_{+})^2-(1+\alpha_{-})^2}{2\Gamma_s^*\sqrt { 2 \pi}}
\left(\frac{\tan^{-1}\sqrt{\xi}}
{\sqrt{\xi}} -\frac { 1 } { 1+\xi } \right)
\end{equation}
where $\xi=\frac{mL^2}{4I}$.
 Comparison of the exact expression of the Independent Kick model with the  simulation results of the BL equation
in Fig.~\ref{fig:velocity_fric6} shows that it has the correct dimensionless dependence
$1/\Gamma_s^*$.

Finally, the regular part of the AVD can be obtained from the inverse Fourier transform of
the characteristic function $\langle e^{i k\Omega} \rangle$. For a Gaussian  bath distribution, by using the change of variable $z=2x/L$, 
one obtains 
\begin{figure}[t]
\resizebox{7.2cm}{!}{\includegraphics{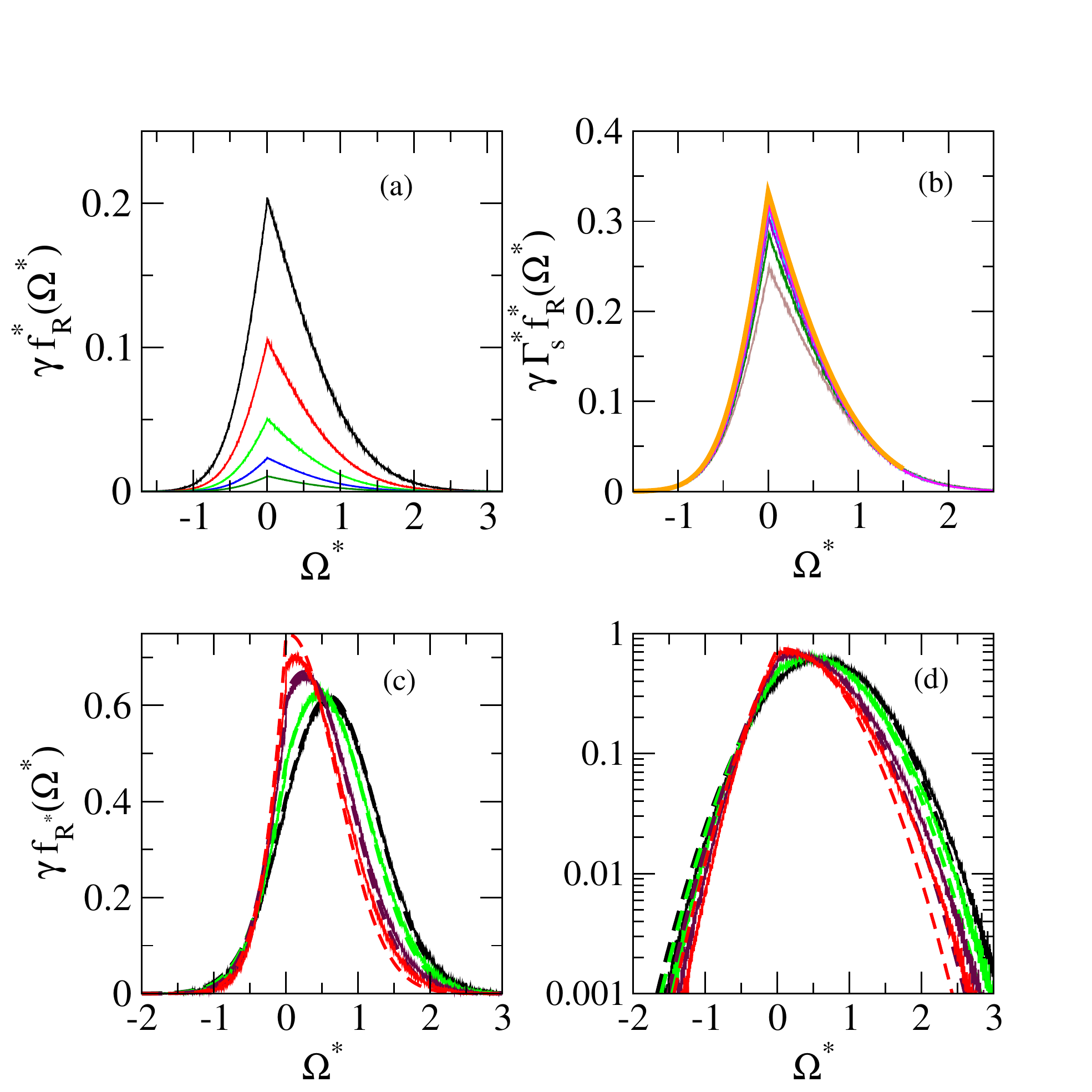}}
 \caption{(Color online) (a) Regular part of the AVD, $\gamma f_R^*(\Omega^*)$,  for various frictional torques
$\Gamma_s^*=1.22,2.72,6.05,13.6,30.0$ (From top to bottom) obtained from numerical simulations of the BL equation.
 $M/m=10$ 
with other parameters the same as in Fig. \ref{fig:velocity_fric6}. 
(b) rescaled $\Gamma_s^*\gamma f_R^*(\Omega^*)$ for the same 
values of the friction torque (from bottom to top). The upper curve corresponds to Eq.\ref{eq:distribution}.
(c) AVD $f_R^*(\Omega^*)$ (full curves) and FP solutions (Eq.\ref{eq:FPF}, dashed curves) for $M/m=20$ and 
$\Gamma_s^*=0.01,0.033,0.074,0.1$. (d) Log-linear plot of $f_R^*(\Omega^*)$ for the same 
values of the friction torque.}\label{fig:distribM10}
\end{figure}

\begin{align}\label{eq:distribution}
f_R^*(\Omega^*)&=\frac{1}{4\sqrt{2\pi}\xi\gamma\Gamma_s^*}\left[\theta(\Omega^*)\int^{1}_{0}
dz \,e^{-\frac {(1+\xi z^2)^2 \Omega^{*2}}{8\xi^2(1+\alpha_{+})^2  z^2}}
\right.
\nonumber\\&\left.+\theta(-\Omega^*)\int^{1}_{0}dz\, e^{-\frac {
(1+\xi z^2)^2 \Omega^{*2}}{8\xi^2(1+\alpha_{-})^2  z^2}}\right]
\end{align}
Therefore, the Independent Kick model provides non-trivial angular velocity distributions which are the sum of a regular function with a non-Gaussian shape,  even in 
the Brownian limit,  with an amplitude decreasing as $1/\Gamma_s^*$ and a singular contribution. 
  Fig.~\ref{fig:distribM10}~a displays regular angular velocity distributions 
of the BL model   for different 
frictional torques. As expected the amplitude decreases as the friction increases. 
Fig.~\ref{fig:distribM10} b  shows  the rescaled distributions  $\Gamma_s^*\gamma f_R^*(\Omega^*)$ versus $\Omega^*$. One observes  
 that for $\Gamma_s^*>2$, the curves converge to the AVD of the Independent Kick model, Eq.~(\ref{eq:distribution}) confirming that it
describes very accurately all kinetic properties of the BL model for moderate to large friction. 

The amplitude of the singular contribution of $f^*(\Omega^*)$ is given 
by $1-\gamma=1-\langle e^{ik\Omega^*}\rangle_{k=0}=1-\frac{
(2+\alpha_++\alpha_-)}{8\xi\Gamma_s^*}\ln\left(1+\xi\right)$.
Consequently, the model remains valid if $\gamma<1$, namely if the frictional
torque is sufficiently large. For moderate friction, the Independent Kick model
overestimates the mean angular velocity given by the BL
equation (see Fig.~\ref{fig:velocity_fric6}).

{\em Brownian Limit and the Fokker-Planck Equation}.
We now consider the opposite limit when the stopping time $\tau_s$ is much larger than 
the intercollision time $\tau_c$ and when $M>>m$. For homogeneous granular motors where the mean angular velocity goes to zero in the Brownian limit
 \cite{Brey1999,Cleuren2007}, the Fokker-Planck can be derived by performing a Kramers-Moyal (KM) expansion of the BL
integral operator. For heterogeneous granular motors, where the mean angular velocity remains finite,  we introduced a KM-like method providing 
the exact Brownian limit\cite{Talbot2011}, but finite mass corrections cannot be easily included. We propose here an
approach that consists of performing a complete series expansion in terms of derivatives of $f(\Omega)$
and a complete resummation of the coefficients allowing the BL operator to be expressed as  
\begin{equation}\label{eq:BLo}
 J[f,\phi]=\sum_{n=1}^\infty
\frac{1}{n!I^n}\frac{\partial^n(g_n(\Omega)f(\Omega))}{\partial \Omega^n}
\end{equation}
with $g_n(\Omega)=\left.\frac{\partial^n (g(\Omega,a)}{\partial
a^n}\right|_{a=0}$ where the generating function $g(\Omega,a)$ is given by 

\begin{align}
 &g(\Omega,a)=2\rho\int_0^{L/2}dx \int_0^\infty dy y
\left[\exp\left(\frac{-(1+\alpha_{+})Imxya}{I+mx^2}\right)\right.\nonumber\\
& \phi(x\Omega
+y)\left. +\exp\left(\frac{(1+\alpha_{-})Imxya}{I+mx^2}\right)\phi(x\Omega
-y)\right]
\end{align}
Truncating the BL operator, Eq. (\ref{eq:BLo}), at  second-order and
adding the frictional torque leads to the following Fokker-Planck equation
\begin{align}\label{eq:FP}
\frac{\partial f(\Omega,t)}{\partial t}&=\frac{1}{I}
\frac{\partial}{\partial
\Omega}\left[(\Gamma_s\sigma(\Omega)+g_1(\Omega)) f(\Omega,t)\right]
\nonumber\\&+\frac{1}{2I^2}\frac{\partial^2}{\partial
\Omega^2}[g_2(\Omega)f(\Omega,t)]=0
\end{align}
in which all finite-mass
corrections are incorporated,  and where deviations from a Gaussian distribution are present for
large finite masses. Recalling  that the $g_n(\Omega)$ are proportional to $\rho$, we see that for a given frictional torque $\Gamma_s$,
increasing the bath density reduces the effect of friction.
The corresponding Langevin equation features a motor torque with a 
non-linear dependence on $\Omega$ and a colored noise.

The steady state solution of Eq. (\ref{eq:FP}) is
\begin{equation}
 f(\Omega)=\frac{C_{\Gamma_s}}{g_2(\Omega)}\exp\left[-2I\int_0^\Omega du\; 
\left(\frac{g_1(u)+\Gamma_s\sigma(u)}{ g_2(u)}\right)\right]
 \end{equation}
where $C_{\Gamma_s}$ is obtained from the normalization condition $\int d\Omega
f(\Omega)=1$. This result clearly shows that, even in the absence of friction, 
the AVD is non-Gaussian for finite mass ratios. 
Moreover, the signature of the presence  of friction is a 
cusp of $f(\Omega)$ at
$\Omega=0$.  This feature, also obtained in the Independent Kick model, is always observed in simulation results of the BL equation.

In the Brownian limit $g_1(\Omega)\sim g'_1(\tilde{\Omega})(\Omega-\tilde{\Omega})$ and
$g_2(\Omega)=2T_g g'_1(\tilde{\Omega})$, where $T_g$ (the rotor granular temperature which is lower than the bath temperature, $T$)
 and $\tilde{\Omega}$ are given by the Kramers-Moyal
expansion (Eqs. 10 and 12 in Ref.~\cite{Talbot2011}). This finally gives a stationary
distribution, at the lowest order in $m/M$,
\begin{equation}\label{eq:FPF}
 f(\Omega)=C \exp\left(-\frac{I(\Omega-\tilde{\Omega})^2}{2T_g}-\frac{I\Gamma_s|\Omega|
} {g_1'(\tilde{\Omega})T_g}\right)
\end{equation}
where $C$ is the normalization constant.
Whereas  one observes a Gaussian  decay of the AVD at large velocity,
$f(\Omega)$ decreases  exponentially for small and intermediate angular velocities, due to  the presence of
friction. The analytical expression (Eq.\ref{eq:FPF}) provides an accurate description of the  AVD 
obtained by simulation (see Fig.\ref{fig:distribM10} c and d).
  A first-order expansion of the dimensionless mean angular velocity can be easily obtained
\begin{equation}\label{eq:OMEGA}
 \langle\Omega^*\rangle= (\alpha_{+}-\alpha_{-})(a_1 -a_2\Gamma_s^*+...)
\end{equation}
where $a_1$ and $a_2$ are positive constants. 
The dominant term corresponds to the first scaling law obtained in simulation results of the BL equation.
More generally, the prediction of the Fokker-Planck equation (dotted curve) is
compared to
simulation results of the
BL equation as  a function of  the frictional torque $\Gamma_s^*$ for
$M/m=10$ (see Fig.~\ref{fig:velocity_fric6}). As expected, the 
Fokker-Planck description is accurate at low and moderate friction, whereas the Independent
Kick model is asymptotically exact in the large fraction limit. Even in the small region of moderate frictions $\Gamma^*$, the two
models slightly overestimate the mean angular velocity. We note that Touchette et al. obtained a time-dependent solution of the FP equation of a particle subject to solid friction and viscous damping \cite{Touchette2010},  
corresponding to   $\alpha_+=\alpha_-$ in the present model. This approach can be extended to obtain full solution of the granular Brownian motor.

In their experimental study\cite{PhysRevLett.104.248001}, 
Eshuis et al. observed that the mean angular velocity increases as $(S-S_{c})^{1.4}$ where 
$S$ is the dimensionless shaking strength and $S_{c}$ is a threshold value. The relation between $S-S_c$ and $T$ is not clear,
 but our analysis 
suggests  two scaling laws could  be observed in experiments corresponding to the small and large friction limits. Further experiments should
 clarify this.

We thank Olivier B\'enichou for helpful discussions.


%

\end{document}